# Spatiotemporal Noise Triggering Infiltrative Tumor Growth under Immune Surveillance

Wei-Rong Zhong[1,2], Yuan-Zhi Shao[1†], Li Li[3], Feng-Hua Wang[3], and Zhen-Hui He[1]

1. Department of Physics, School of Physics and Engineering, Sun Yat-sen University, Guangzhou 510275, P.R.China.
2. Centre for Nonlinear Studies, Hong Kong Baptist University, Kowloon Tong, Hong Kong
3. South-China State-key Laboratory for Oncology, Cancer Center of Sun Yat-Sen University, Guangzhou 510006, P.R.China.

**Abstract**
A spatiotemporal noise is assumed to reflect the environmental fluctuation in a spatially extended tumor system. We introduce firstly the structure factor to reveal the invasive tumor growth quantitatively. The homogenous environment can lead to an expansive growth of the tumor cells, while the inhomogenous environment underlies an infiltrative growth. The different responses of above two cases are separated by a characteristic critical intensity of the spatiotemporal noise. Theoretical and numerical results present a close annotation to the clinical images.

## 1. Introduction

Expansive and infiltrative growths are two major forms of tumor invasion. In general, the former corresponds to benign tumors and the latter is regarded as one of the features of malignant tumors (1). Figure 1 produced by ourselves shows two typical clinical computerized tomographic (CT) pictures of these two growth forms, for example, brain glioma (A region) and meningoma (B region). Apparently, brain meningoma produces an expansive growth pattern with a clear-cut envelope, while brain glioma typically yields another diffuse, infiltrative growth pattern. Tumor cells can sometime change from a benign to a malignant state on a special condition. How does a tumor exhibit a malignant infiltrative growth pattern? To answer this question, some scientists reported their molecular biological annotations, e.g., the influences of the extracellular matrix, genic heterogensis, the podoplanin and the integrins on the invasive tumor growth (2-6). In these annotations a common point is that the changing microenvironment of tumor cells affects them crucially. From the physical point of view, the environmental fluctuation has a main effect on the growth and diffusion of tumor cells.

In previous mathematical models, the emphases on the tumor growth-diffusion using reaction-diffusion equations without fluctuations have yielded many valuable results

† To whom correspondence should be addressed at: Department of Physics, Sun Yat-sen University, Guangzhou 510275, P.R.China. E-mail: stssyz@mail.sysu.edu.cn.

(7-12). Likewise, new insights into tumor growth, for example, the study of noise effects on the tumors, has gained many interesting findings (13-15). These studies, however, have not given a sufficient consideration to the environmental fluctuation under immune surveillance against tumor, and furthermore, how to analyze the growth pattern of tumor invasion and metastasis is less clear.

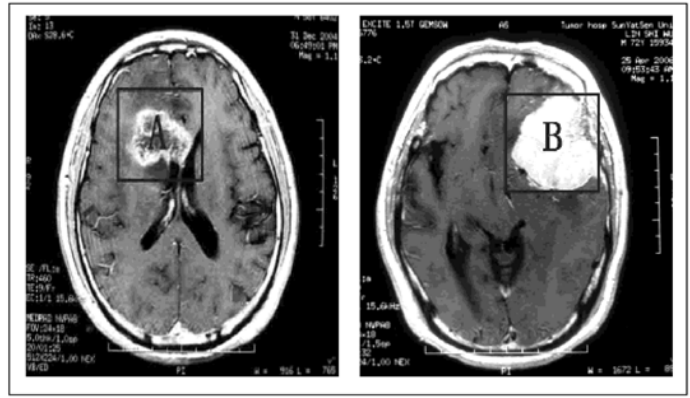

**Fig.1** Clinical CT images of brain glioma (A) and meningoma (B). The squares guide to the regions of tumors. A and B are the infiltrative and the expansive growth, respectively.

In this article, we will model the invasion growth of tumor systems under the inhomogenous microenvironment using two dimensional spatiotemporal stochastic differential equations and give a new insight into the molecular mechanism of the invasive tumor growth. In the current work, we introduce the structure factor to evaluate the order degree of the spatiotemporal structure (ODSS) of the invasive tumor growth; a characteristic critical intensity of the spatiotemporal fluctuation is proposed, which categorizes the tumor invasion into the expansive and infiltrative growth form.

**2. Model and equation**
Considering the growth of tumor cells under an immune surveillance against cancer modeled by the enzyme dynamic process (16,17), we assume that the two variable species diffusion dynamic equations can be expressed as below:

$$\frac{\partial u_i}{\partial t} = r u_i (1 - \frac{u_i}{K}) - \frac{\beta u_i^2}{1 + u_i^2} + D_u (1 - \varepsilon v_i) \nabla^2 u_i,$$

$$\frac{\partial v_i}{\partial t} = \frac{\beta u_i^2}{1 + u_i^2} - \theta v_i + D_v u_i \nabla^2 v_i.$$

(1)

Where $u_i$ and $v_i$ are the densities of tumor cells and dead cells at site $i$, respectively; $r$ is the linear per capita birth rate of tumor cells, and $K$ is the carrying capacity of the environment. $\beta u_i^2 / (1 + u_i^2)$, defined as an immune form, quantifies the abilities of immune cells to recognize and attack tumor cells. $D_u$ and $D_v$ are the diffusion coefficients of $u_i$ and $v_i$, respectively. $\epsilon$ and $\theta$ are the velocity coefficients. Generally, the environmental fluctuation and tumor heterogeneity may lead to a spatiotemporal fluctuation of tumor growth rate $r$ (18), which indicates that the growth rate fluctuates not only in time (evolution fluctuation) but also in space (inhomogenization). Therefore the spatiotemporal fluctuation is a main feature of the tumor system, and the growth rate $r$ in

Eq.(1) should be rewritten as $r_0 + \xi_i(t)$, where $\xi_i(t)$ is the Gaussian noise, white in time and space, with zero mean and autocorrelation defined by $\langle \xi_i(t) \rangle = 0$, $\langle \xi_i(t)\xi_j(t') \rangle = 2\sigma \delta_{i,j}\delta(t-t')$, in which $\sigma$ is the noise level and $i,j$ are lattice sites. Assuming that the dead cells can be decompounded by the environmental tissues rapidly, we write the equivalent single variable stochastic differential equation of Eq.(1) as (16,17),

$$\frac{du_i}{dt} = r_0 u_i(1 - \frac{u_i}{K}) - \frac{\beta u_i^2}{1+u_i^2} + u_i(1 - \frac{u_i}{K})\xi_i(t) - \frac{D}{2d}\sum_{j\in n(i)}(u_i - u_j), \quad (2)$$

here $n(i)$ is the set of the *2d* nearest neighbors of site *i*; *d* and *D* are the spatial dimension and the diffusion coefficient, respectively.

The above equations are general and cover different kinds of tumor growth and diffusion phenomena, especially nonequilibrium growth. We would like to seek the existence of nonequilibrium phase transition induced by multiplicative noise, in systems described by these equations. Such a phase transition is characterized by the appearance of multiple steady state probability distributions $p_{st}(\{u_i\})$, which has been applied successfully in numerous stochastic problems (19,20).

If set $f(u_i) = r_0 u_i(1-u_i/K) - \beta u_i^2/(1+u_i^2)$ and $g(u_i) = u_i(1-u_i/K)$, one will obtain the corresponding Fokker-Planck equation of Eq.(2),

$$\frac{\partial p(\{u_i\},t)}{\partial t} = -\frac{\partial [A(u_i)p(\{u_i\},t)]}{\partial u_i} + \frac{\partial^2[B(u_i)p(\{u_i\},t)]}{\partial u_i^2}, \quad (3)$$

in which

$$A(u_i) = f(u_i) + \sigma g(u_i)g'(u_i) - \frac{D}{2d}\sum_{j\in n(i)}(u_i - u_j), \quad (4)$$

$$B(u_i) = \sigma g^2(u_i).$$

In Ref.(18), we have detailed the method of the calculation. For simplicity of notation, here we drop the subscript *i*. The stationary solution to the Fokker-Planck equation of Eq.(2) is given to be,

$$p_{st}(u) = Z\exp[\frac{1}{\sigma}\int_0^u dv \frac{f(v) - \sigma g(v)g'(v) - D[v - E(v)]}{g^2(v)}], \quad (5)$$

where *Z* is a normalization constant,

$$f(v) = r_0 v(1-v/K) - \beta v^2/(1+v^2),$$

$$g(v) = \sigma v(1-v/K), \quad (6)$$

and

$$E(v) = \langle v_i | v_j \rangle = \int v_j p_{st}(v_j | v_i) dv_j, \quad (7)$$

which represents the steady state conditional average of $v_j$ at neighboring sites $j\in n(i)$, given the value $v_i$ at site *i*.

The maxima of $p_{st}(u)$ are obtained from $f(u) - \sigma g(u)g'(u) - D[v - E(v)] = 0$. For the low value of $\sigma$, figure 2 illustrates the SPD produces a monostable state. With increasing noise intensity, the SPD changes from a monostable state to a bistable state. Note that the spatiotemporal noise can change the potential of the tumor growth and diffusion. This kind of bistable state easily leads the system to two separate phases in space (19,20), which agrees with the following numerical simulations.

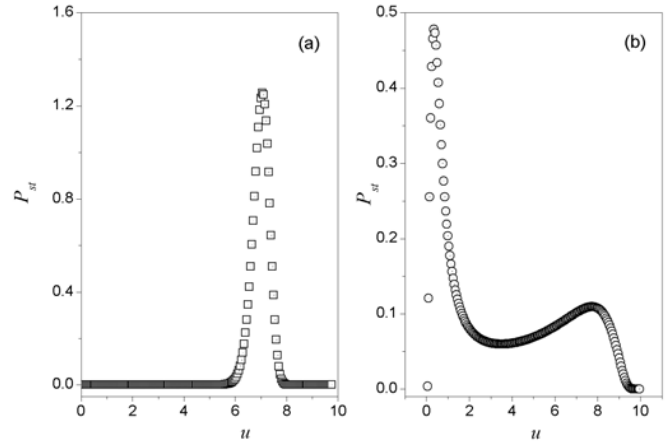

**Fig.2** Stationary probability distributions of average population of tumor cells for different noise intensities $\sigma=0.01$ (a), 0.40 (b). The parameters are $r_0=1.0$, $\beta=2.12$, $K=10.0$, and $D=0.5$.

## 3. Result and analysis

We performed numerical computations of Eq. (1) using the two dimensional (2D) cellular automaton method (21,22). We set $r_0=1.0$, $\beta=2.2$, $K=10.0$, and $D=0.8$, respectively, and configure the initial condition as a 2D Gaussian distribution function with the maximum in the center of the lattice. To be closer to the clinical observation, we carried out 2D computations systematically at various spatiotemporal noise levels. Figure 3 displays six typical evolution patterns for the duration of 0, 900, and 2700 in the case of the low and the high value of noise intensity $\sigma$, respectively. The tumor invasion exhibits an expansive growth form for the low value of $\sigma$, while it undergoes an infiltrative growth form for the high value of $\sigma$. The tumors with the expansive growth feature grow up to a definite size and stop expanding with a clear boundary. The tumor cells with infiltrative growth will, however, spread and overgrow through the whole lattice rapidly.

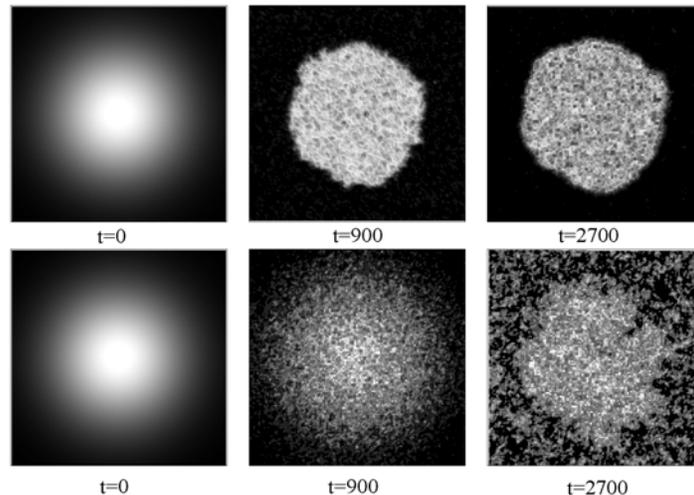

**Fig.3** Evolution patterns of tumor invasion using cellular automatons of Eq. (2) with periodic boundary at times 0, 900, and 2700. The top three pictures are for the noise level $\sigma=0.15$ and the bottom three pictures are corresponding to the noise level σ=0.60. They have the same initial condition. The system size is 128.

In order to analyze these patterns more quantitatively, we defined the structure factor (SF) [which is obtain by circularly averaging the normalized form factor $S(k_x, k_y, t)$] (19,20)

$$S(k,t) = \langle u(k,t)u(-k,t) \rangle \qquad (8)$$

where $u(k, t)$ is the spatial Fourier transform of the autocorrelation density of the cells $u_i(k, t)$; $k$ is the magnitude of the wave vector **k**, and $t$ the time. Usually, a sharp peak in $S(k, t)$ discloses the existence of an order spatiotemporal structure. Figure 4 shows that, in the case of a low noise level, the peak of the SF increases with the time, i.e., the ODSS rises during the tumor invasion. Conversely, a descendant trend is observed for the high noise level. Judging by the SF of the clinical images shown in Fig.5, one can find that the expansive growth produces a sharp peak while the infiltrative growth exhibits a low peak of SF. Here all clinical images have been processed using the same gray scale criterion, and the SF of the clinical images is also obtained through Eq.(4). This suggests that the tumor with expansive growth be a system just as the conservative system in physics, which is less influenced by the environmental fluctuations due to the rounded envelope and clear boundary. Besides the brain glioma and meningoma mentioned above, we studied also giant cell tumor and osteogenic sarcoma of bone, which produce also the same evidence of Figs.1 and 4. Therefore it is probably a universal phenomenon that expansive growth corresponds to benign tumors, while infiltrative growth to malignant tumors.

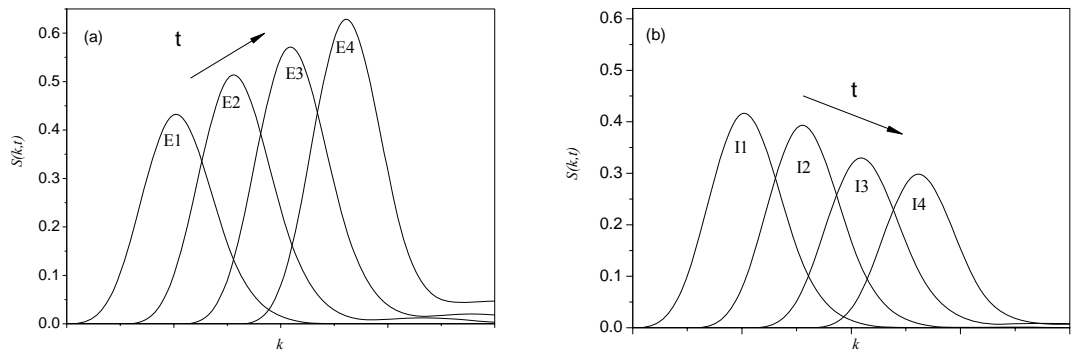

**Fig.4** Graphs of the structure factor S(k,t) for σ=0.15 (a) and 0.60 (b) at various times t. The times of the left figure are 30, 240, 1200, and 3000, respectively. The times of the right figure are 60, 300, 1200, and 3000, respectively.

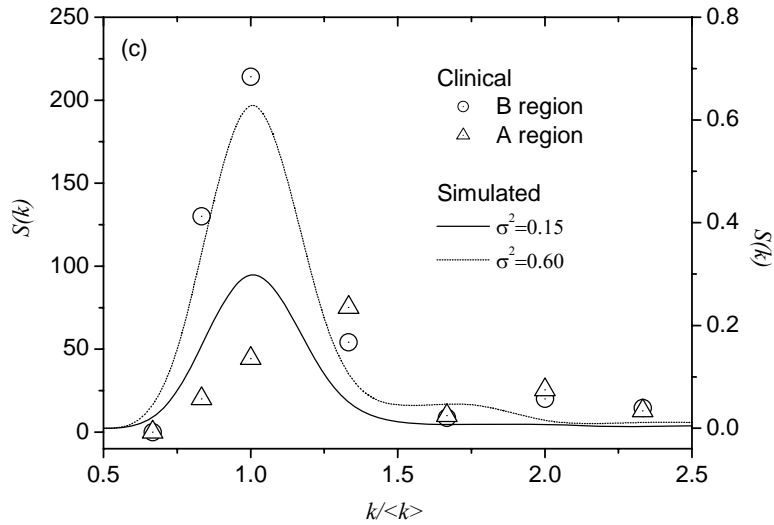

**Fig.5** The structure factors of the expansive and the infiltrative growth. The circles and triangles are clinical results corresponding to the square regions (A and B) of Fig.1, respectively. The solid and dot line are simulated results.

The sharpness of the spatiotemporal structure can be also quantified by considering the height of the SF over its half-peak width, which is denoted by $P(t)$. This quantity, as shown in Fig.6(a), varies at first and then tends to a stable value. Different trends are observed for different noise levels, namely, ascending trend for low noise level and descending one for high counterpart. Calculating the variance $\Delta P$ between initial and stable values of $P(t)$, we show in Fig.6(b) that $\Delta P$ has a linear relationship with the square of noise levels. If $\Delta P$ is negative, it means a decrease of the ODSS of the tumor growth pattern during the tumor invasion. Conversely, if $\Delta P$ is positive, it means a increase. The critical noise level, at which $\Delta P$ equals zero, classifies the tumor invasion to two growth types, i.e., expansive growth under weak spatiotemporal fluctuations and infiltrative growth under strong spatiotemporal fluctuations.

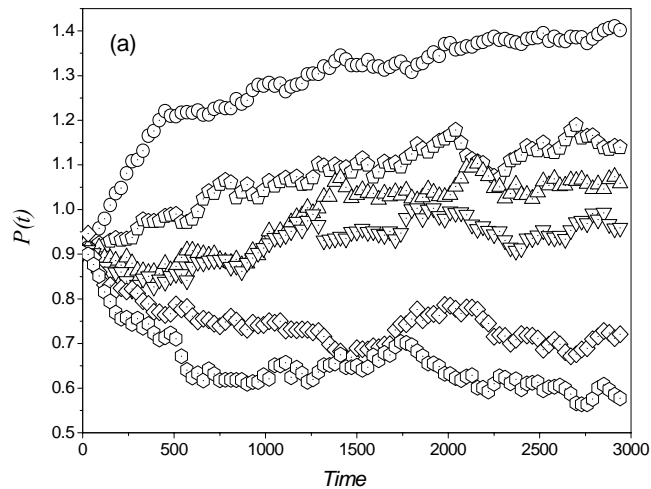

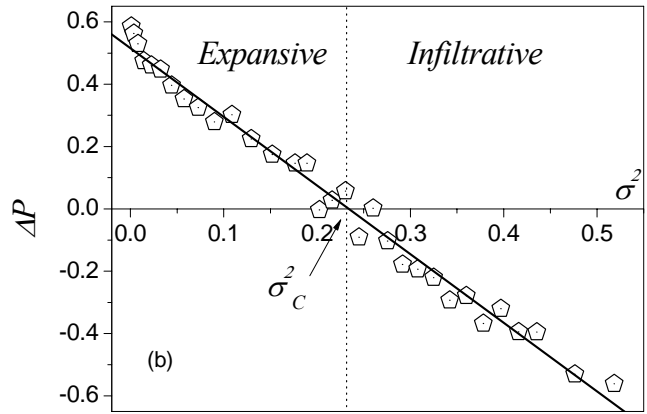

**Fig.6** (a) Evolution of the quantity P(t) for the noise levels σ=0.15, 0.36, 0.42, 0.48, 0.54, and 0.60 (from top to bottom). (b) $\Delta P$ as a linear function of the square of the noise level $\sigma^2$. Here $\Delta P$ is defined by $\oint P(t)dt/T - P(0)$, where $P(0)$ is the initial value of $P(t)$ and $T=5000$. The scatters are corresponding to numerical results. The solid line is a fitting solution, $\Delta P = 0.515 - 2.200\sigma^2$, $\sigma_c^2 = 0.234$.

We have to take into account an additional point when comparing theoretical and simulated patterns with clinical images. The infiltrative growth patterns of our model fill the whole plane and the denser inner proliferative region is located in the center of the patterns. In clinical images, however, the denser region looks like a loop with a dark center for infiltrative growth. This little contrast probably contributes to the dead cells that have been assumed to be decompounded immediately in our model. Nevertheless, the invasive growth forms and the structure factors of our simulations are in principle consistent with the clinical data.

## 4. Conclusion

In summary, the clinical images show that the growth forms are quite different for benign and malignant tumor cells. The former takes on an expansive growth while the latter an infiltrative growth. We have firstly introduced a spatiotemporal noise into the tumor system to model its environmental fluctuations and formulated the structure factor to quantitatively characterize the evolution patterns of tumor invasion. We suggest the benign tumors with expansive growth be regarded as a system with weak fluctuation and high spatiotemporal order, and the malignant tumors as the one with a strongly fluctuating, nonconservative environment. The change in the order of the spatiotemporal structure of a tumor has a linear relationship with the square of the noise level. There exists a critical noise level to distinguish the expansive from the infiltrative growth form.

**Acknowledgements**
We would particularly like to thank Prof. Collins at Boston University for his valuable comments and encouragements. This work was partially supported by the National Natural Science Foundation (Grant No. 60471023), People's Republic of China.